\documentclass[epj]{svjour}
\usepackage{epsfig}
\newcommand{\average}[1]{\left\langle{#1}\right\rangle}
\newcommand{\vs}{\textit{vs.}}

\newcommand{\eg}{\textit{e.g.}}
\renewcommand{\Im}{\mathop\mathrm{Im}}
\renewcommand{\Re}{\mathop\mathrm{Re}}
\newcommand{\tw}{{t_\mathrm{w}}}
\newcommand{\teff}{{T^\mathrm{eff}}}
\newcommand{\tmes}{{T^\mathrm{meas}}}
\newcommand{\tmeas}{\tmes}

\begin{document}

\title{Measuring effective temperatures in out-of-equilibrium
driven systems}
\titlerunning{Measuring effective temperatures}

\author{Rapha{\"e}l Exartier\inst{1,2}\fnmsep\thanks{E-mail:
{\tt exartier@ccr.jussieu.fr}}
\and
Luca Peliti\inst{2,3}\fnmsep\thanks{E-mail: {\tt peliti@na.infn.it}}
\fnmsep\thanks{Associato INFN, Sezione di Napoli (Italy).}}

\institute{Laboratoire des Milieux D{\'e}sordonn{\'e}s et
H{\'e}t{\'e}rog{\`e}nes\thanks{Laboratoire associ\'{e} au
CNRS (URA n$^{\circ}$ 800) et \`{a} l'Universit\'{e}
Pierre-et-Marie Curie--Paris VI.}\\
Tour 13 -- Case 86, 4 place Jussieu, F--75252 Paris Cedex 05 (France)
 \and Dipartimento di Scienze Fisiche and Unit{\`a} INFM,
Universit{\`a} ``Federico II''\\
Mostra d'Oltremare, Pad.~19, I--80125 Napoli (Italy)
\and
Laboratoire de Physico-Chimie Th{\'e}orique\\
Ecole Sup{\'e}rieure de Physique et Chimie Industrielles\\
10, rue Vauquelin, F--75231 Paris Cedex 05 (France)}
\date{Received: date / Revised version: date}
\mail{%
Rapha{\"e}l Exartier.}

\abstract{We introduce and solve a model of a thermometric
measurement on a driven glassy system in a stationary state. We
show that a thermometer with a sufficiently slow response measures
a temperature higher than that of the environment, but that the
measured temperature does not usually coincide with the effective
temperature related to the violation of the
Fluctuation-Dissipation Theorem.
} 
\maketitle
\section{Introduction}
\label{intro}
Thermal equilibrium is a rather subtle concept. It
relies on the distinction between ``fast" and ``slow"  processes
with respect to a given  macroscopic measurement. It follows that
the same system can be at equilibrium on one scale and out of
equilibrium on another. More strikingly it can be at equilibrium
but exhibiting different properties on two scales at
once~\cite{Ma}.

The notion most intimately connected to equilibrium is
temperature. It is operationally defined by the so-called zeroth
law of thermodynamics, which states that when two systems are in
thermal equilibrium with a third one, then they must be in thermal
equilibrium with each other. This allows one to define temperature
as a signature of the equivalence class defined by mutual thermal
equilibrium. This property makes possible the use of test systems,
called ``thermometers", to decide whether any two systems will or
will not remain in thermal equilibrium when brought into contact.
When two systems are not in mutual equilibrium, the direction of
the energy flow between them is determined by the second law of
thermodynamics.

When dealing with non-equilibrium systems the challenge is thus to
produce an ``effective" time-scale dependent temperature that would
predict the direction of heat flows within this scale.

Indeed, in the context of weak turbulence, Hohenberg and Shraiman
\cite{HS} have defined an effective temperature for stationary
non-equilibrium systems through an expression involving the
response, the correlation and the temperature of the heat
reservoir. A closely related expression appears in the theory of
aging systems~\cite{cukuglass}. In a recent work \cite{cukupe},
these views have been unified for a class of out-of-equilibrium
systems with small heat flows, which includes nonstationary pure
relaxational systems, like glasses, and stationary systems, slowly
driven by non-re\-lax\-a\-tion\-al for\-ces. The concept of
effective temperatures has been further reviewed in
refs.~\cite{cuku,laeticia}. In a recent experiment temperatures
higher than the thermal bath temperatures have been exhibited in
an oscillating circuit coupled to an aging glycerol sample after a
quench~\cite{Grigera}.

In the present work we analyze the process of a thermometric
measurement in a glassy system, by means of an exactly solvable
model. We restrict ourselves to the stationary
non-e\-qui\-li\-brium regime of a driven system. We consider a
simple system and a thermometer, both described by Langevin
equations. By taking advantage of Time-Translation Invariance
(TTI) the Langevin equation is transformed into an algebraic
equation in Fourier space, that can be analytically solved.

Our thermometer is a simple physical system coupled to its own
heat bath, which is different from the thermal bath of the
observed system. We suppose that during the measuring time, the
two systems are brought into contact, each being coupled with its
own thermal bath. We then monitor the exchanged energy between the
system and the thermometer in the stationary regime. The reading
of the thermometer corresponds to the temperature of the heat bath
of the thermometer for which the net energy flow between the
system and the thermometer vanishes. We discuss the relation of
the measured temperature with the effective temperature defined in
refs.~\cite{HS,cukupe}.

In section 2 we recall the generalization of the
Fluc\-tu\-a\-tion-Dis\-si\-pa\-tion Theorem to nonequilibrium
systems and show how it defines an effective  temperature. In
section 3 we describe the general measurement procedure. In
section 4 we specify the procedure for the measurement of the
effective temperature of an asymmetric spherical SK model.
Finally, section 5 is devoted to the analysis of the results
obtained for this system.

\section{Fluctuation-Dissipation Theorem and effective temperatures}
According to~\cite{cukupe}, the definition of an effective
temperature for non-equilibrium systems can be related to the
violation of the Fluctuation-Dissipation Theorem (FDT). Let us
consider a system (described by the Hamiltonian $H$) subject to a
time-dependent perturbation of the form:
\begin{equation}
H\longrightarrow H-h(t)O,
\end{equation}
where $O$ is an extensive operator. The correlation function $C(t,t')$
of $O$ is defined by
\begin{eqnarray}
C(t,t')&:=&\average{O(t)O(t-)}_\mathrm{c}\nonumber\\
&:=&\average{O(t)O(t')}-\average{O(t)}\average{O(t')},
\end{eqnarray}
while the corresponding response function $R(t,t')$ is given by
\begin{equation}
R(t,t'):=\left.\frac{\delta \average{O(t)}}{\delta h(t')}
\right|_{h\equiv 0}.
\end{equation}
For a system at equilibrium with a thermal reservoir at
temperature $T$, Time-Translation Invariance (TTI) intimates that
both the correlation and the response functions depend only on the
time difference $\tau$ between their time arguments ($\tau=t-t'$).
On the other hand, the FDT entails a relation between the response
and the correlation functions:
\begin{equation}
R(\tau)={\theta(t-t')\over T}{\partial\over\partial t'}
\average{O(t)O(t')}_\mathrm{c}
=-{\theta(\tau)\over T}{\partial C(\tau)\over\partial \tau}.
\end{equation}
Experimentally, one usually measures the (time-)in\-te\-gr\-a\-ted
susceptibility:
\begin{equation}
\chi(\tau):=\int^\tau_0d\tau'\, R(\tau').
\end{equation}
At equilibrium, we can use the FDT to compare this susceptibility
to the correlation function:
\begin{equation}
\chi(\tau)={1-C(\tau)\over T}.
\end{equation}
(We are considering magnetic systems for which $C(0)=1$.) Then, a
parametric plot of $\chi(\tau)$ \vs\ $C(\tau)$ yields a straight
line with slope equal to $-1/T$.

A certain class of out of equilibrium systems with very slow
dynamics exhibits an aging regime in which the FDT is violated in
a very specific way. Following an initial quench of temperature,
these systems fall out of equilibrium and do not reach it again,
even on macroscopic time scales. The longer the time $\tw$ elapsed since
the initial quench, the slower is the response of the
system to a given perturbation: the system ages. This phenomenon
also appears in the time-correlation functions. In these systems,
even in the limit $\tw\to\infty$, a parametric plot of
$\chi(t,\tw)$ \vs\ $C(t,\tw)$ does not yield a straight line with
slope $-1/T$  as in equilibrium.

\begin{figure}[htb]
\begin{center}
\epsfig{file=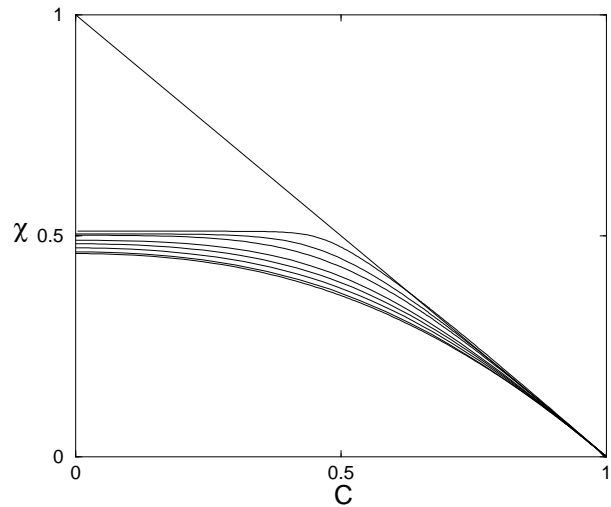,width=8cm}
\end{center}
\caption{Plot of $\chi$ \vs\ C of the asymmetric spherical SK
model for $T_1=10$, $J=20$ and different values of the asymmetry
parameter $v$. The susceptibility $\chi$ is normalized by the bath
temperature $T_1$. The lines correspond (from above to below) to
$v=0.2;0.3;0.4;0.5;0.6;0.7;0.8$. } \label{fig:CKplots}
\end{figure}

Very similar features appear in some \textit{stationary}
non-equilibrium (driven) systems
\cite{cukuglass,cukupe,cuku,laeticia,Horner}, like the one shown
in figure~\ref{fig:CKplots}.
Let us introduce a parameter $v$  which measures the intensity of
the driving  force. (For these systems, of course, the time needed
to reach the stationary nonequilibrium state diverges as $v\to0$.)
In a sense, $v$ plays a role similar to $\tw$ in aging systems:
the smaller $v$, the ``older'' the system. In a driven system TTI
is satisfied, but FDT is not, even in the limit of vanishing
driving force ($v\to 0$)~\cite{cukupe}. Let us consider the slope
$\chi'(C)=d\chi/dC$ of the curve $\chi(C)$. According to
\cite{cukupe}, for small enough driving forces, the
\textit{effective temperature} $\teff$ can be expressed in terms
of this slope:
\begin{equation}
\label{def:Teff}
\teff(C):=-\frac{1}{\chi'(C)}=-\left({d\chi(C)\over dC}\right)^{-1}.
\end{equation}In all known cases
one has $\teff(C)\geq T$.

We can define a time scale $\tau(q,v)$ by means of the relation:
\begin{equation}
\label{timescale}
C(\tau(q,v),v)=q.
\end{equation}
If $q$ is larger than a threshold value $q_{\rm EA}$
(called the Ed\-wards-An\-der\-son order parameter) one has
\[\lim_{v\to 0}\tau(q,v)<\infty.\] This is equivalent to the following
definition of $q_{\rm EA}$:
\begin{equation}
q_{\rm EA}:=\lim_{\tau\to \infty}\lim_{v\to 0}C(\tau,v).
\end{equation}
On the other hand, if $q<q_{\rm EA}$, the time $\tau(v)$
diverges as $v$ goes to zero.

Let us thus consider a thermometric measurement, performed
on a characteristic time scale $\tau$. We wish to compare
it with $\teff(\tau)$, where
\begin{equation}
\teff(\tau):=-\left.\frac{1}{\chi'(q)}\right|_{q=C(\tau,v)}.
\label{def:teff(tau)}
\end{equation}

\section{Measurement procedure}
The measurement procedure is similar to the one described in
Appendix C of \cite{cukupe}. It does not crucially depend on the
nature of the thermometer, as long as it satisfies the
fluctuation-dissipation theorem and has a tunable response time.
We use a small but macroscopic thermometer in contact with a
thermal bath at temperature $T_2$. The driven system whose
effective temperature is to be measured is in contact with a bath
at temperature $T_1$.

The thermometer is coupled to the observable $O_1(S_1)$ of the
system via its observable $O_2(S_2)$. The interaction Hamiltonian
writes: $H_{\rm int}=-a\,O_1(S_1)O_2(S_2)$. We remark that $O_1$
and $O_2$ are conjugate to each other. After a certain time, which
depends on the coupling constant $a$ and on $|T_1-T_2|$, a
stationary regime appears.

We define as $\tmes$ the value of the temperature $T_2$ for which
the net energy transfer between the system and the thermometer
vanishes at stationarity. This temperature is compared with the
effective temperature $\teff$ of the system, defined by
eq.~(\ref{def:Teff}).

During the measurement procedure both the system and the
thermometer are kept in contact with their own baths. The
resulting system has two temperatures. (A simple system with two
temperatures has been introduced and discussed in
ref.~\cite{expe}.)

At stationarity, the net  power  gain for the thermometer,
$\dot Q_2$, writes:
\begin{eqnarray}
\dot Q_2&=&a\,\langle{\dot O_1 O_2}\rangle=
\lim_{\tau\to0} a\,\partial_{\tau}\tilde C_{12}(\tau),
\end{eqnarray}
where we have introduced the cross correlation function
\begin{equation}
\label {C12}
\tilde C_{12}(\tau):=
\lim_{t\to\infty}\average{O_1(t+\tau)O_2(t)}_\mathrm{c}.
\end{equation}
Using linear response, one has
\begin{eqnarray}
O_1(t)=O_{1\mathrm b}(t)+a\int_0^t dt\,
R_{1}(t-t') O_{2\mathrm b}(t'),\\
O_2(t)=O_{2\mathrm b}(t)+a\int_0^t dt\,
R_{2}(t-t') O_{1\mathrm b}(t'),
\end{eqnarray}
where $O_{i\mathrm b}$ are the observables in the absence of
coupling and $R_{i}$ the corresponding response function
($i=1,2$). At stationarity, to first order in the coupling $a$,
the cross correlation function (\ref{C12}) is given by
\begin{eqnarray}
\tilde C_{12}(\tau)&=&a\int_{-\infty}^\tau d\tau' R_{1}(\tau-\tau')
C_2(\tau')  \nonumber \\
&&{}+a\int_{-\infty}^0 d\tau'R_2(-\tau') C_1(\tau'-\tau),
\end{eqnarray}
where we have introduced the correlation functions of the
\textit{bare} systems:
\begin{equation}
C_i(\tau):=\lim_{t\to\infty}\langle O_{i\mathrm b}(t+\tau)
O_{i\mathrm b}(t)\rangle_\mathrm{c}.
\end{equation}
Then the rate of heat transfer writes:
\begin{equation}
\dot Q_2 =a^2\int_0^\infty d\tau \left( R_2(\tau)\partial_\tau
C_1(\tau)-R_1(\tau)\partial_\tau C_2(\tau)\right).
\end{equation}
From eq.~(\ref{timescale}), we can substitute $q$ for $\tau$ as
the integration variable: $d q=\dot C_1(\tau)\,d\tau$. 
One can now exploit the fluctuation dissipation relations
for the
bare systems, namely
$R_1(\tau)=-\dot C_1(\tau)/\teff(\tau)$
and $R_2(\tau)=-\dot C_2(\tau)/T_2$, and obtain
\begin{equation}
\label{dotQ_2}
\dot Q_2=a^2\int_0^1 dq \,R_2(q) 
\left ( \frac{T_2}{\teff(q)}-1 \right ). 
\end{equation}

The measured temperature $\tmes$ is defined as the one which makes
$\dot Q_2$ to vanish:
\begin{equation}
\label{Tmeas}
\tmeas(q_2)^{-1}:=\frac{ \int_0^1 dq R_2(q,q_2)\teff(q)^{-1} }
{ \int_0^1 d\tau R_2(q,q_2)}.
\end{equation}
We have introduced the parameter $q_2:=C_1(\tau_2)$, where $\tau_2$
represents the tunable  characteristic time of the thermometer. 
$\tmeas(q_2)^{-1}$ is the average of $\teff(q)^{-1}$ weighted
by $R_2(q,q_2)$.
We remark that the measured temperature is independent of the
coupling constant $a$, provided that it is small enough to ensure
the validity of the linear response theory.
On the other hand, $\tmeas$ depends strongly on $q_2$,
because the lower boundary of the integrals appearing 
in eq.~(\ref{Tmeas}) is effectively cutoff at $q_2$.

For a system at equilibrium $\teff(q)=T_1$ for any $q$.
Thus $\tmeas=T_1$ as expected, whatever the characteristic
time of the thermometer.

Let us consider a simple system with only two time sectors:
\begin{equation}
\teff(\tau)= \cases{
T_1, &for $ \tau \leq \tau_{\rm EA} $;\cr
T_1'>T_1, &for $\tau \geq \tau_{\rm EA}$.}
\end{equation}

\begin{figure}[htb] 
\begin{center} 
\epsfig{file=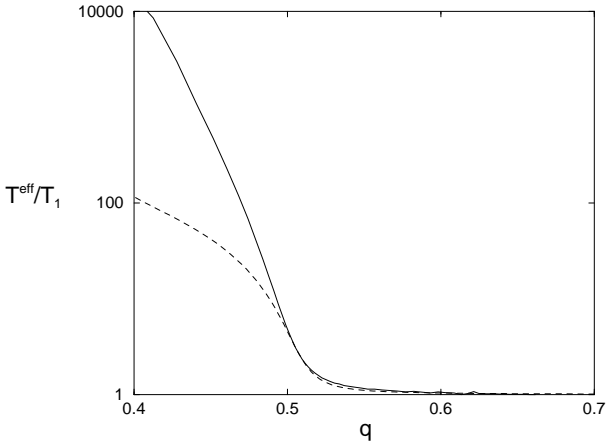,width=8cm} 
\end{center} 
\caption{Comparison of the effective and measured temperatures of the 
asymmetric spherical SK model for $T_1=10$,
 $J=20$ and $v=.1$. The solid line
corresponds to the effective temperature of the bare model. The dashed 
line corresponds to the termometer measure. 
\label{fig:Tmeas_Teff-q-v.1} 
}
\end{figure}

We have introduced the notation $\tau_{\rm EA}$ defined by the
relation: $C_1(\tau_\mathrm{EA})=q_{\rm EA}$.
For $q_2 \geq q_{\rm EA}$, which corresponds to probing 
the short time behaviour of the aging system, the lower boundary
cutoff at $q_2$ of the integrals of eq.~(\ref{Tmeas}) implies that the
effective temperature of the driven system is constantly $\teff(q)=T_1$ 
over the integration interval and thus: 
\begin{equation}
\tmeas(q_2)= T_1.
\end{equation}
For $q_2 \leq q_{\rm EA}$  the 
temperature $\tmeas$ is not equal to $\teff=T_1'$.
 Splitting the numerator of
(\ref{Tmeas}) in two integrals from $0$ to $q_{\rm EA}$ and from
$q_{\rm EA}$ to $1$ we obtain 
\begin{eqnarray}
\label{Tmeas-aging}
\tmeas(q_2)^{-1} &=&\frac{1}{T_1'} 
\frac{ \int_0^{q_{\rm EA}}dq\, R_2(q,q_2)}{ 
\int_0^1 d\tau\, R_2(q,q_2)}\nonumber\\
&&{}+ \frac{1}{T_1}\frac{\int_{q_{\rm EA}}^1dq\, R_2(q,q_2)}{ 
\int_0^1 d\tau\, R_2(q,q_2)} .
\end{eqnarray}
The measured inverse temperature is a weighted average of the inverse 
temperature of the 
bath $1/T_1$ and the inverse effective temperature $1/\teff$. Its value
is intermediate between them, as shown
in figure \ref{fig:Tmeas_Teff-q-v.1}, where the aging system 
and the thermometer are the ones described in the next section.

\begin{figure}[htb] 
\begin{center} 
\epsfig{file=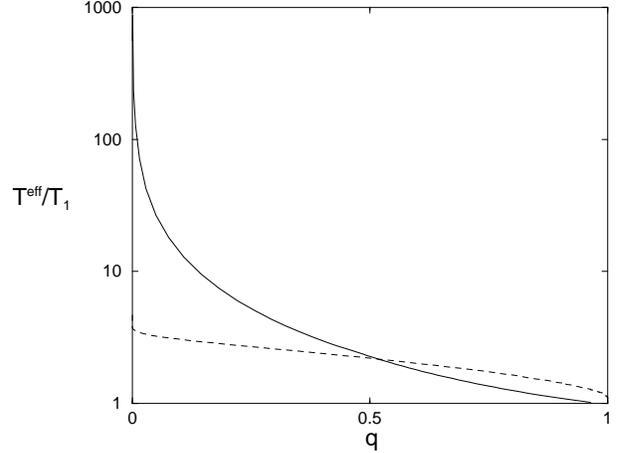,width=8cm} 
\end{center} 
\caption{Comparison of the effective and measured temperatures of the 
asymmetric spherical SK model for $T_1=10$,
 $J=20$ and $v=.9$. The solid line
corresponds to the effective temperature of the bare model. The dashed 
line corresponds to the termometer measure. 
\label{fig:Tmeas_Teff-q-v.9} 
}    
\end{figure}

If we consider a driven system with many times scales,
the measured temperature over a certain time scale $q_2$ is the 
weighted average of all the effective temperatures of the system over
this time scale. Since, on one hand, $R_2$ is a decreasing function of
$q$ and, on the other hand, $\teff$ is an increasing function of $q$,
we expect  the measured temperature to be \textit{lower} than the 
effective temperature when the thermometer probes the long time scales
corresponding to the limit $q\to0$. Nevertheless, for intermediate 
time scales, if $R_2(q,q_2)$ is not peaked sharply enough around $q_2$,
it is possible 
to measure a temperature higher than the effective one, as  
shown, \eg, in 
figure \ref{fig:Tmeas_Teff-q-v.9} .

\section{Effective temperature of a spherical SK model with randomly
asymmetric bonds}
The previous considerations can be made more explicit in
an exactly solvable model of a system-thermometer complex.

We consider a spherical SK model with randomly asymmetric
bonds~\cite{Crisanti}. The Hamiltonian has the form
\begin{equation}
H_1:=-\frac{1}{2}\sum_{i,j=1}^NJ^{\rm s}_{ij}S_1^iS_1^j
+\frac{r_1}{2}\sum_{i=1}^{N}(S_1^{i})^2.
\end{equation}
The ``spin'' variables $S_1^i$ can take any real value. The
parameter $r_1$ is a Lagrange multiplier which enforces the
spherical constraint $\sum_{i=1}^{N}(S_1^{i})^2=N$. The
interaction matrix $J^{\rm s}_{ij}$ is a symmetric matrix whose
diagonal elements vanish and whose off-diagonal elements, for each
pair $\{i,j\}$ of indices, are independent Gaussian variables with
zero mean and the following variance:
\begin{equation}\label{variance:eq}
\overline{(J_{ij}^{\rm s})^2}={J^2\over N}{1\over 1+v^2}.
\end{equation}
The parameter $v>0$ is a measure of the strength of the driving
force (see below). This spherical SK model is coupled to a
single-spin paramagnetic thermometer via a bilinear interaction of
strength $a$. The Hamiltonian of the paramagnet is given by
\begin{equation}
H_2:=\frac{r_2}{2} (S_2)^2.
\end{equation}
The response time scale of the paramagnet is given by
$\tau_2:=1/r_2$. Indeed, we recall that the bare response function
of a paramagnet has the expression
\begin{equation}
R_2(t)=\theta(t)\exp^{-r_2t}.
\end{equation}
The total Hamiltonian writes
\begin{equation}
{H}:={H}_1(S_1) +{H}_2(S_2) +H_\mathrm{int}(S_1,S_2),
\end{equation}
where the interaction Hamiltonian $H_\mathrm{int}$
is given by
\begin{equation}
H_\mathrm{int}:=-a\,S_2\sum_{i=1}^N S_1^i.\label{def:Hint}
\end{equation}
Stability requires $a²<r_1r_2$. The dynamics of the system is
described by a system of linear Langevin equations:
\begin{eqnarray}\label{sys:1,S,S,glass,para}
\partial_t S^i_1(t)&=&-\frac{\partial{H}}{\partial S^i_1}
+b_i(S_1)+\eta^i_1(t), \\
\partial_t S_2(t)&=&-\frac{\partial{H}}{\partial S_2}+\eta_2(t).
\end{eqnarray}
In eq.~(\ref{sys:1,S,S,glass,para}), the \textit{driving field}
$b_i(S_1)$ is given by:
\begin{equation}
b_i(S_1):=vJ^{\rm as}_{ij}S_1^j,
\end{equation}
where $J_{ij}^{\rm as}$ is an antisymmetric matrix whose
off-diagonal elements, for each pair $\{i,j\}$ of indices, are
independent Gaussian random variables of zero mean and variance
equal to that of $J^{\rm s}_{ij}$ (eq.~(\ref{variance:eq})). The
$\eta^i_1$  are thermal noises at temperature $T_1$
 with zero mean and variance given by
$\langle \eta^i_1(t) \eta^i_1(t')\rangle=2 T_1 \delta(t-t')$,
while $\eta_2$ is a thermal noise at temperature $T_2$.

In the thermodynamical limit ($N\to\infty$), it is possible to
average over the disorder by the means of dynamical functional
integration techniques. Thus the equations for the asymmetric
spherical SK model in (\ref{sys:1,S,S,glass,para}) reduce to a
single equation for a single spin $S_1$. The new system of
equations reads
\begin{eqnarray}                \label{sys:2,S,S,glass,para}
\partial_t S_1(t)&=&-r_1(t) S_1(t) +a S_2(t)\\
&&{}+ J'^2\int_{t_0}^t dt'\, R_{11}(t,t')S_1(t') + \eta_1(t),
\nonumber\\
\partial_t S_2(t)&=&-r_2 S_2(t) +a S_1(t) + \eta_2(t),
\end{eqnarray}
where $J'=\sqrt{(1-v^2)/(1+v^2)}\,J$, and $\eta_1$ is a
renormalized Gaussian thermal noise with zero mean and variance
given by
\begin{equation}
\langle \eta_1(t) \eta_1(t')\rangle=2 T_1 \delta(t-t')
+ J^2 C_{11}(t,t').
\end{equation}

For the response we obtain the following autonomous system:
\begin{eqnarray}
\label{sys:R,glass,para}
(\partial_t+r_1(t))R_{11}(t,t')&=&aR_{21}(t,t')+\delta(t-t')\\
&&{}+J^{\prime2}\int_{t_0}^t dt''R_{11}(t,t'')R_{11}(t'',t'),
\nonumber\\
(\partial_t+r_1(t))R_{12}(t,t')&=&aR_{22}(t,t')\\
&&{}+J^{\prime2}\int_{t_0}^t dt''R_{11}(t,t'')R_{12}(t'',t'),
\nonumber\\
(\partial_t+r_2)R_{22}(t,t')&=&aR_{12}(t,t')+\delta(t-t') ,\\
(\partial_t+r_2)R_{21}(t,t')&=&aR_{11}(t,t').
\label{sys1:R,glass,para}
\end{eqnarray}
The time $t_0$ can be freely chosen between the time the interaction
was switched on and the observation times $t$ and $t'$.
The equations for the correlation function involve the response:
\begin{eqnarray}
\label{sys:C,glass,para}
&&(\partial_t+r_1)C_{11}(t,t')= aC_{21}(t,t')+2T_1R_{11}(t',t)\\
&&\qquad {}+\int_{t_0}^t dt'' \left(J^{\prime2}R_{11}(t,t'')
C_{11}(t'',t')\right.\nonumber\\
&&\qquad\qquad\qquad\left.{}+J^2C_{11}(t,t'')R_{11}(t',t'')\right),
\nonumber\\
&&(\partial_t+r_1)C_{12}(t,t')= aC_{22}(t,t')+2T_1R_{21}(t',t)\\
&&\qquad{}+\int_{t_0}^t dt''
\left(J^{\prime2}R_{11}(t,t'')C_{12}(t'',t')\right.\nonumber\\
&&\qquad\qquad\qquad\left.{}+J^2C_{11}(t,t'')R_{21}(t',t'')\right),
\nonumber\\
&&(\partial_t+r_2)C_{22}(t,t')= aC_{12}(t,t')+ 2T_2R_{22}(t',t),\\
\label{sys1:C,glass,para}
&&(\partial_t+r_2)C_{21}(t,t')= aC_{11}(t,t')+ 2T_2R_{12}(t',t).
\end{eqnarray}
After some time, the system enters a stationary regime where
$C_{ij}(t,t')=\hat C_{ij}(t-t')$ and $R_{ij}(t,t')=\hat
R_{ij}(t-t')$. Choosing $t_0$, $t'$ and $t$ in this regime, and
taking advantage of Fourier analysis, one can solve the system
for the response and then the one for the correlation. For
simplicity we suppose that the interaction between the system and
the thermometer has been switched on an infinite time in the past.
This corresponds to sending $t_0\to-\infty$. All the quantities
depend on the value of $r_1$ which is chosen in order to verify
the spherical condition: $C_{11}(t,t)={1\over N}\sum_{i=1}^N
{S_1^i}^2=1$. Taking the derivative of this condition with respect
to time we obtain
\begin{equation}
\lim_{t'\to t^{-}} {\partial C_{11}(t,t')\over \partial t}
+\lim_{t'\to t^{+}}{\partial C_{11}(t,t')\over \partial t}=0.
\end{equation}
Substituting eq.~(\ref{sys:C,glass,para}) we
obtain an equation for $r_1$:
\begin{eqnarray}
\label{eq:r1,glass,para}
r_1&=&T + a C_{21}(0) \\
&&{}+ (J^2+J^{\prime2})\int_{t_0}^t dt''\,
R_{11}(t,t'')C_{11}(t'',t').\nonumber
\end{eqnarray}
Since $C_{21}$, $R_{11}$ and $C_{11}$ all depend on $r_1$ this is an
equation for $r_1$. We solved it numerically and then substituted the
value of $r_1$ into the correlation and response functions. After this
step the solution is completed and we can search for the temperature
$T_2$ of the paramagnetic thermometer which makes the heat flux to
vanish.

The power exchanged between the thermometer and the system at
stationarity is given by
\begin{equation}
\dot Q_2:=\average{\frac{\partial H_\mathrm{int}}{\partial S_1}
\dot S_1}-\average{\frac{\partial H_\mathrm{int}}{\partial S_2}
\dot S_2}.
\end{equation}
From the definition (\ref{def:Hint}) of $H_\mathrm{int}$ we obtain
\begin{eqnarray}
\dot Q_2(t)&=&-aS_2(t){d S_1(t)\over d t}+aS_1(t){d S_2(t)\over d t}
 \\
&=&\lim_{t'\to t}\left(-a\,\partial_t C_{21}(t,t')
+a\partial_tC_{12}(t,t')
\right). \nonumber
\end{eqnarray}
By using the results of the appendix and taking an inverse Fourier
transformation, we finally obtain
\begin{eqnarray}
\dot  Q_2=a \int {d\omega\over 2\pi}\,i\omega
\,\left(\tilde C_{12}(\omega)-\tilde C_{21}(\omega)\right)\nonumber\\
\label{Q1}
=-2 a \int {d\omega\over 2\pi}\,\omega\,
\Im\left(\tilde C_{12}(\omega)\right).
\end{eqnarray}
The temperature measured by the thermometer is the one which
makes the heat flux $\dot Q_2$ to vanish. The resulting measured
temperatures are shown in figure \ref{fig:Tmeas-q-v} and
should be compared with the expected effective temperature
shown in figure \ref{fig:Teff-q-v}.

\section{Results}
\begin{figure}[htb]
\begin{center}
\epsfig{file=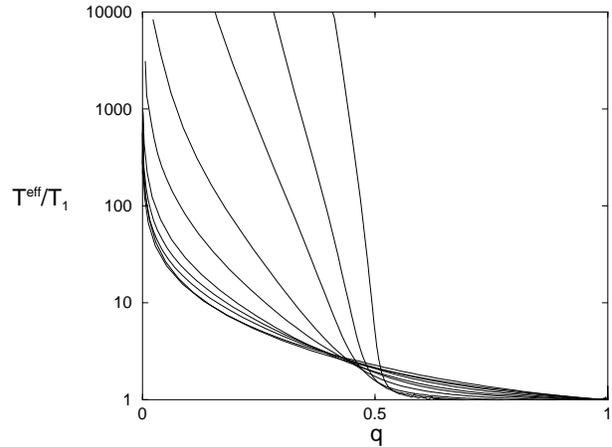,width=8cm}
\end{center}
\caption{Effective temperature $\teff$ \vs\ $q$ for the bare
asymmetric spherical SK model for $T_1=10$, $J=20$, and for
different values of the asymmetry parameter $v$. $\teff$ is
normalized by the temperature $T_1$ of the thermal bath coupled to
the aging system. The lines correspond (from above to below) to
$v=0.1;0.15;0.2;0.3;0.4;0.5;0.6;0.7;0.8;0.9$. The Edwards-Anderson
order parameter $q_{EA}$ of the corresponding symmetric model is
given by $q_{EA}=T_1/J=.5$, and corresponds to the value of $q$
from where the curves diverge.} \label{fig:Teff-q-v}
\end{figure}

\begin{figure}[htb]
\begin{center}
\epsfig{file=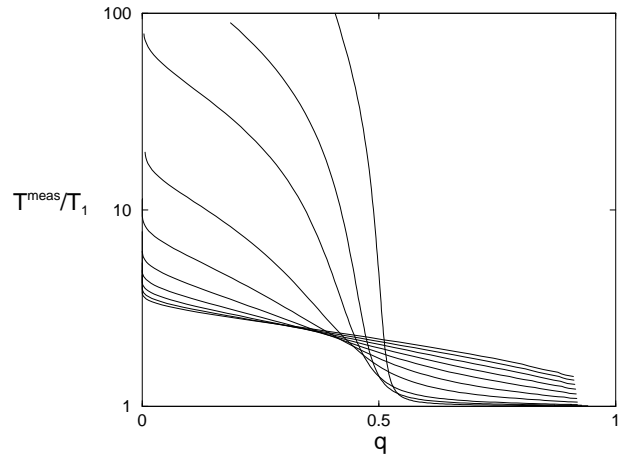,width=8cm}
\end{center}
\caption{Measured effective temperature $\tmes$ \vs\ $q$
of the asymmetric spherical SK model for $T_1=10$, $J=20$, $a=.1$, and
 for different values of the asymmetry $v$. The lines correspond
(from above to below) to $v=0.1;0.15;0.2;0.3;0.4;0.5;0.6;0.7;0.8;0.9$.}
\label{fig:Tmeas-q-v}
\end{figure}

\begin{figure}[htb]
\begin{center}
\epsfig{file=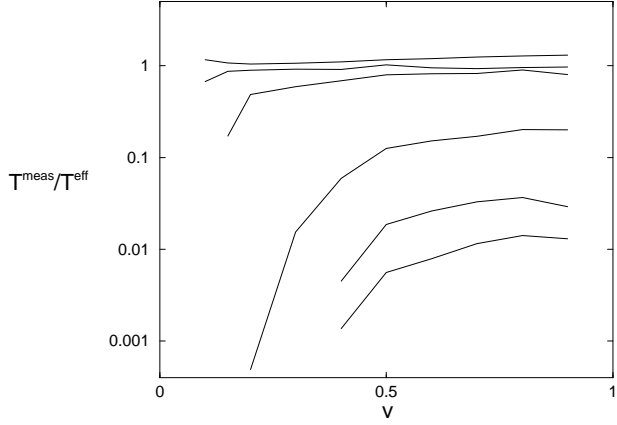,width=8cm}
\end{center}
\caption{Ratio $\tmes/\teff$ \vs\ asymmetry
parameter $v$ of the asymmetric spherical SK model for $T_1=10$,
$J=20$, and for different values of the overlap $q$.
The lines correspond (from above to below) to
$q=0.8;0.5;0.4;0.1;0.01;0.001$.}
\label{fig:TmeasTeff-v-q}
\end{figure}

Figures \ref{fig:Teff-q-v} and \ref{fig:Tmeas-q-v} show
the \textit{effective} and \textit{measured} values of
the temperature as a function of the value of the correlation
function, for the model under study. The two quantitites
behave similarly, in that they are close to the equilibrium
value $T_1$ for $q>q_\mathrm{EA}$, and start increasing,
as $q$ becomes smaller and smaller, for
$q<q_\mathrm{EA}$. Nevertheless it is possible to
identify a \textit{quantitative} discrepancy, analogous
to the one discussed in section 2. For small values of the asymmetry
parameter $v$, the driven system
exhibits two clearly separated regimes, with $\teff=T_1$ for 
$q>q_{\rm EA}$, and a temperature increase for $q<q_{\rm EA}$. For
high values of $q$, $\tmeas$ 
remains close to $T_1$, but is much smaller than $\teff$ in the low-$q$
region,  as shown in fig.~\ref{fig:Tmeas_Teff-q-v.1}. For higher 
$v$,  $\teff$ increases smoothly all along the range of $q$. Being an 
average of $\teff$ over a given time scale, $\tmeas$  becomes quickly 
higher than $T_1$ as it feels the increasing of $\teff$ even for short 
time scales.
Then it keeps on increasing, at a slower pace than $\teff$, as shown
in fig.~\ref{fig:Tmeas_Teff-q-v.9}.  
Figures \ref{fig:TmeasTeff-v-q} show the ratio $\tmeas/\teff$
as a function of the asymmetry parameter $v$. 
The ratio remains close to the ideal value 1
only for $q>q_\mathrm{EA}$. Ideally,
similar plots would apply to an aging system as a
function of the inverse waiting time, because of the
correspondence between $v$ and the waiting time discussed above.

\begin{figure}[htb]
\begin{center}
\epsfig{file=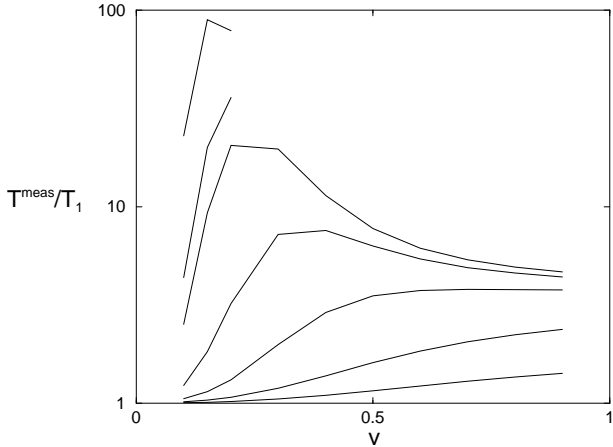,width=8cm}
\end{center}
\caption{Measured effective temperature $\tmes$ \vs\ asymmetry
parameter $v$ of the asymmetric spherical SK model for $T_1=10$, $J=20$
, $a=.1$, and for different values of the parameter $\tau_2$ which sets
the characteristic time of the thermometer. $\tmes$ is normalized by the
 temperature $T_1$ of the thermal bath coupled to the aging system.
The lines correspond (from above to below) to
$\tau_2=100;10;1.0;0.1;0.01$.
}
\label{fig:Tmes-v-r}
\end{figure}

\begin{figure}[htb]
\begin{center}
\epsfig{file=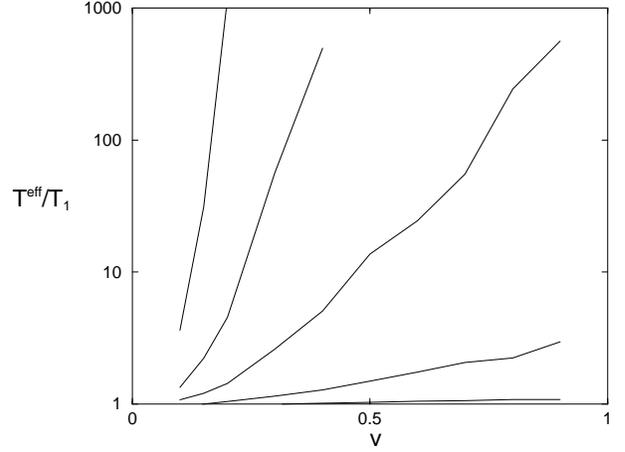,width=8cm}
\end{center}
\caption{Effective temperature $\teff$ \vs\ asymmetry
parameter $v$ of the asymmetric spherical SK model for $T_1=10$, $J=20$
and for different values of $\tau$.
$\teff$ is normalized by the temperature $T_1$ of the
thermal bath coupled to the aging system. The lines correspond
(from above to below) to $\tau=100;10.;1.;0.1;0.01$.}
\label{fig:Teff-v-r}
\end{figure}

In figure \ref{fig:Tmes-v-r} and \ref{fig:Teff-v-r}
the behavior of the measured and of the
effective temperature is shown as a function of the asymmetry parameter
$v$ is shown for different values of the thermometer
response time $\tau_2$. The figures are in qualitative agreement only 
for short time scales which are represented by the lower lines
of the plots. When the thermometer probes longer times scales,
$\tmes(\tau_2)$ shows a non
monotonic behaviour which does not appear in the $\teff(\tau)$ plot.
In the waiting time representation, a thermometer with
a fixed (but long) reaction time $\tau_2$ would first yield higher
and higher readings, as it attempts to approach the flat part of
the graph shown in fig.~\ref{fig:CKplots}, but will
eventually read the temperature of the thermal bath as
$\tw\gg \tau_2$. This leads to a non monotonic behavior
as a function of $\tw$, and an analogous one as a function of
$v$, at least for sufficiently slow thermometers.

In conclusion we have shown in an exactly solvable model
of a thermometric measurement in a ``glassy'' system that, while
thermometer do indeed measure temperatures higher than the
one of the environment if they are slow enough, the relation
of the measured temperature with the effective temperature
defined, \eg, in~\cite{cukupe} is far from trivial.

\begin{acknowledgement}
LP acknowledges the support of a Chaire Joliot of the ESPCI.
Discussions with Serge Galam are gratefully acknowledged.
\end{acknowledgement}

\appendix

\section*{Appendix}

Taking the Fourier transform of the autonomous system for response functions,
equations (\ref{sys:R,glass,para}--\ref{sys1:R,glass,para}), we obtain
\begin{eqnarray}
\label{sys:w,R,glass,para,1}
(i\omega+r_1)\tilde R_{11}(\omega)&=& a\tilde R_{21}(\omega)
+J^{\prime2}\tilde R_{11}(\omega)^2+ 1,\\
\label{sys:w,R,glass,para,2}
(i\omega+r_1)\tilde R_{12}(\omega)&=& a\tilde R_{22}(\omega)
+J^{\prime2}\tilde R_{11}(\omega)\tilde R_{12}(\omega),\\
\label{sys:w,R,glass,para,3}
(i\omega+r_2)\tilde R_{22}(\omega)&=& a\tilde R_{12}(\omega)+ 1,\\
\label{sys:w,R,glass,para,4}
(i\omega+r_2)\tilde R_{21}(\omega)&=& a\tilde R_{11}(\omega).
\end{eqnarray}
Again, from (\ref{sys:C,glass,para}--\ref{sys1:C,glass,para})
we obtain the following equation for the correlation functions:
\begin{eqnarray}
\label{sys:w,C,glass,para,1}
(i\omega+r_1)\tilde C_{11}(\omega)&=&a\tilde C_{21}(\omega)
+J^{\prime2}\tilde R_{11}(\omega)\tilde C_{11}(\omega)\\
&&+J^2\tilde C_{11}(\omega)\overline{\tilde R_{11}(\omega)}
+2T_1\overline{\tilde R_{11}(\omega)},\nonumber\\
\label{sys:w,C,glass,para,2}
(i\omega+r_1)\tilde C_{12}(\omega)&=&a\tilde C_{22}(\omega)
+J^{\prime2}\tilde R_{11}(\omega)\tilde C_{12}(\omega)\\
&&+J^2\tilde C_{11}(\omega)\overline{\tilde R_{21}(\omega)}
+2T_1\overline{\tilde R_{21}(\omega)},\nonumber\\
\label{sys:w,C,glass,para,3}
(i\omega+r_2)\tilde C_{22}(\omega)&=&a\tilde C_{12}(\omega)
+2T_2\overline{\tilde R_{22}(\omega)},\\
\label{sys:w,C,glass,para,4}
(i\omega+r_2)\tilde C_{21}(\omega)&=&a\tilde C_{11}(\omega)
+2T_2\overline{\tilde R_{12}(\omega)}.
\end{eqnarray}
In Fourier space, the equation (\ref{eq:r1,glass,para}) for the
spherical  parameter $r_1$ writes:
\begin{eqnarray}
\label{eq:w,r1,glass,para}
r_1(t)&=&T + a \int {d\omega\over2\pi} \tilde C_{21}(\omega)\nonumber\\
&&{}+ (J^2+J^{\prime2})\int {d\omega\over2\pi} \tilde R_{11}(\omega)
\tilde C_{11}(\omega).
\end{eqnarray}
Equations (\ref{sys:w,R,glass,para,1}--\ref{eq:w,r1,glass,para})
form a nine-equation system for the response and correlation
functions and for the spherical parameter, which is possible to solve
explicitly.
Equations (\ref{sys:w,R,glass,para,1}) and
(\ref{sys:w,R,glass,para,2})  yield a second-order algebraic equation
for $\tilde R_{11}(\omega)$:
\begin{equation}
\label{eq:w,R11,glass,para,1}
J^{\prime2} \tilde R_{11}^2-\left(i\omega+r_1-\frac{a^2}{i\omega+r_2}
\right)\tilde R_{11}+1=0 .
\end{equation}
We choose the solution which respects the symmetries of 
$\Im(\tilde R_{11}(\omega))$, $\Re(\tilde R_{11}(\omega))$, and
recovers the right value for $\tilde R_{11}(\omega)$ in the limit
$a\to0$.

From equations (\ref{sys:w,R,glass,para,2}) and
(\ref{sys:w,R,glass,para,3}) we obtain
\begin{equation}
\label{eq:w,R12,R21,glass,para}
\tilde R_{12}(\omega)=\tilde R_{21}(\omega)
=\frac{a}{i\omega+r_2}\tilde R_{11}(\omega),
\end{equation}
while equation (\ref{sys:w,R,glass,para,3}) yields:
\begin{equation}
\label{eq:w,R22,glass,para}
\tilde R_{22}(\omega)=\frac{1}{i\omega+r_2}+\frac{a^2}
{(i\omega+r_2)^2}\tilde R_{11}(\omega).
\end{equation}
From equations (\ref{sys:w,C,glass,para,1}) and
(\ref{sys:w,C,glass,para,4}) we obtain:
\begin{eqnarray}
\tilde C_{11}&=&\frac{2T_1 + 2T_2 {a^2}/{(\omega^2
+r_2^2)}}{|\tilde R_{11}|^{-2}-J^2},\\
\tilde C_{21}&=&\frac{a}{\omega^2+r_2^2}\left [
(r_2-i\omega)\tilde C_{11}+2T_2\overline {\tilde R_{11}(\omega)}
\right].
\end{eqnarray}
As shown by equation (\ref{Q1}), the measurement procedure of $\tmes$
gives a particular relevance to the imaginary part of
$\tilde C_{21}(\omega)$:
\begin{equation}
\Im(\tilde C_{21})=-a\frac{2T_2\Im(\tilde R_{11})
+\omega\tilde C_{11}}{\omega^2+r_2^2}
\end{equation}
\end{document}